%% file: main.tex
\documentclass[a4paper, 10pt, conference]{ieeeconf}      % Use this line for a4
\usepackage{hyperref}
\usepackage{listings}
\usepackage{subcaption}
\usepackage{graphicx}
\title{\LARGE \bf
Automatic Generation of a Hybrid Query Execution Engine %for SQLite
}

\author{ 
	\parbox{3 in}{
    	\centering Aleksei Kashuba\\
        Vrije Universiteit Amsterdam\\
        Amsterdam, The Netherlands\\
        {\tt\small a.kashuba@student.vu.nl}
    }
	\hspace*{ 0.5 in}
	\parbox{3 in}{
    	\centering Hannes M{\"u}hleisen\\
        Centrum Wiskunde \& Informatica\\
        Amsterdam, The Netherlands\\
        {\tt\small hannes@cwi.nl}
     }
}

\begin{document}

\maketitle
\thispagestyle{empty}
\pagestyle{empty}

%%%%%%%%%%%%%%%%%%%%%%%%%%%%%%%%%%%%%%%%%%%%%%%%%%%%%%%%%%%%%%%%%%%%%%%%%%%%%%%%
\begin{abstract}

The ever-increasing need for fast data processing demands new methods for efficient query execution. Just-in-time query compilation techniques have been demonstrated to improve performance in a set of analytical tasks significantly. In this work, we investigate the possibility of adding this approach to existing database solutions and the benefits it provides. 
To that end, we create a set of automated tools to create a runtime code generation engine and integrate such an engine into SQLite which is one of the most popular relational databases in the world and is used in a large variety of contexts. Speedups of up to 1.7x were observed in microbenchmarks with queries involving a large number of operations.

\end{abstract}

%%%%%%%%%%%%%%%%%%%%%%%%%%%%%%%%%%%%%%%%%%%%%%%%%%%%%%%%%%%%%%%%%%%%%%%%%%%%%%%%

\section{INTRODUCTION}

Query processing is an essential part of any database management system. The traditional way of executing a SQL query is to convert it into a query plan consisting of relational algebra operators, which communicate with each other through an iterator interface. It is a simple and extensible design which allows for various optimizations at the query plan level and became known as Volcano-style query processing \cite{graefe1993volcano}. However, this method was designed when I/O performance was the main issue, and it is therefore not optimized for CPU performance. In the last decade, faster and cheaper main memory has made the Volcano-style query execution model less attractive for modern database management systems, which prompted the development of new optimizations for query processing one of which is called just-in-time compilation. 

The fundamental principle of JIT compilation is to generate code on the fly once the query is known and then compile it into native machine instructions. Using the information available at runtime allows to tailor the algebraic operators for the specific query and eliminate the interpretation overhead. Additional improvements come from the optimizations performed by the compiler \cite{viglas2014just}. The method has proved to be effective and was adopted in both commercial (Hekaton \cite{freedman2014compilation}) and research projects (HIQUE\cite{krikellas2010generating}, HyPer \cite{neumann2011efficiently}, Peloton \cite{menon2017relaxed}). One of the approaches to JIT compilation named ``holistic query evaluation'' is to generate a source file implementing the functionality of the query using a set of templates, compile the code and dynamically link it with the database server \cite{krikellas2010generating}.

The idea of runtime code generation for query execution is not new and was part of IBM System R developed in 1970's where a procedure to execute the query was assembled from components consisting from machine language instructions \cite{chamberlin1981history}. Still, the successor of System R did not utilize the technique because of the impracticality of maintaining a code generation engine for a machine language, e.g., porting it to a different operating system \cite{viglas2014just}. Modern compilers, such as LLVM, use an intermediate representation as a lightweight abstraction over machine code. Writing a code generator for an IR is simpler than writing a code generator for a machine language but is still time-consuming and challenging to maintain. Writing a query interpreter is more straightforward, portable and allows to reduce latency for short running queries if a cost model estimates the compilation time to be too high. Therefore, we investigate the possibility of automatically deriving a code generation engine from the interpreter. Such an approach could reduce the workload associated with maintaining a code generation engine while still providing significant performance improvements.  

The contributions of this paper are the following.

\begin{itemize}
\item We present a template-based code generation engine for the popular embedded DBMS SQLite \cite{sqlite} and describe a method for automatic extraction of operator templates directly from the source code of the query interpreter. 

\item We design a microbenchmark and a set of experiments to investigate the performance of the JIT query execution engine in relation to the complexity of a query and its selectivity. The new system provides a speedup of up to 1.7x in complex queries.

 \item The JIT system is compared to the standard SQLite as well as a second baseline that implements a technique for the reduction of the interpretation overhead without compilation called ``direct threading''. 

\end{itemize}

The rest of the paper is organized as follows. Section 2 summarizes the relevant parts of the internal organization of SQLite and the direct threading technique. Section 3 describes how the JIT system functions and the algorithm for automatic template generation. Section 4 presents the performed experiments and their results. Section 5 reviews relevant work. Section 6 describes future work and concludes.

%The rest of the paper is organized as follows: TODO

% A performance evaluation of the proposed system was conducted by comparing it to the standard version of SQLite as well as a modified one which implements a different technique used for the reduction of interpreter overhead commonly known as direct threading. Direct threading is a popular optimization in programming languages interpreters and makes use of computed gotos to reduce the number of executed instructions and potentially improve branch prediction. The technique is further discussed in the experimental study section.

\begin{figure}[thpb]
\centering
\begin{lstlisting}[numbers=left,language=C]
int sqlite3VdbeExec(
  Vdbe *p              // The VDBE 
){
  // The array of operations
  Op *aOp = p->aOp; 
  // Current operation
  Op *pOp = aOp;
  ...
  for(pOp=&aOp[p->pc]; 1; pOp++){
      switch( pOp->opcode ){
        case OP_Goto: { 
        jump_to_p2:
          pOp = &aOp[pOp->p2 - 1];
          break;
        }
        case OP_Ge: {
          ...
          break;
        }
     }
   }
}
\end{lstlisting}
\caption{Opcode interpreter (simplified)}
\label{switch}
\end{figure}

\section{BACKGROUND INFORMATION}
    
SQLite somewhat departs from the traditional Volcano-style query execution. Instead of linking algebraic operators between each other with indirect function calls it creates a query execution plan in the form of its custom bytecode. The bytecode program and other necessary information such as open cursors (pointers into a table) are contained in an instance of a ``Virtual Database Engine'' (VDBE) \cite{VDBE}. The interpreter contains a \texttt{switch} statement which determines the code that needs to be executed based on the opcode of the current instruction. In the remainder of the paper, we refer to the code that is executed when a particular case is encountered as a ``case block''.

 Each SQLite instruction consists of an opcode and up to three operands. The first and the third operands usually contain the addresses of virtual registers, the contents of which should be processed. The second operand is usually the next program counter for a jump.  For each produced result a callback function is invoked, and the processing of the opcode continues until the end of the program. 

Figure \ref{switch} shows the simplified source code of the interpreter function in SQLite. For each invocation of the function, the pointer to the correct operation is set at the beginning of the for-loop by accessing the array of opcodes at the position specified by the virtual program counter (Line 7). Then the opcode is decoded in the \texttt{switch} statement, and the case block corresponding to it is executed. After a \texttt{break} statement in the case block, the pointer to the current operation is incremented to point to the next opcode. To move execution of the bytecode to a different point in the program, opcodes perform a jump to \texttt{jump\_to\_p2} label, where the current operation pointer is set to point to the opcode at the position stored in \texttt{p2}. 
\begin{figure}[thpb]
\centering
\begin{lstlisting}[language=SQL]
  Query: SELECT i FROM test WHERE i<20;
\end{lstlisting}
\includegraphics[width=0.5\textwidth]{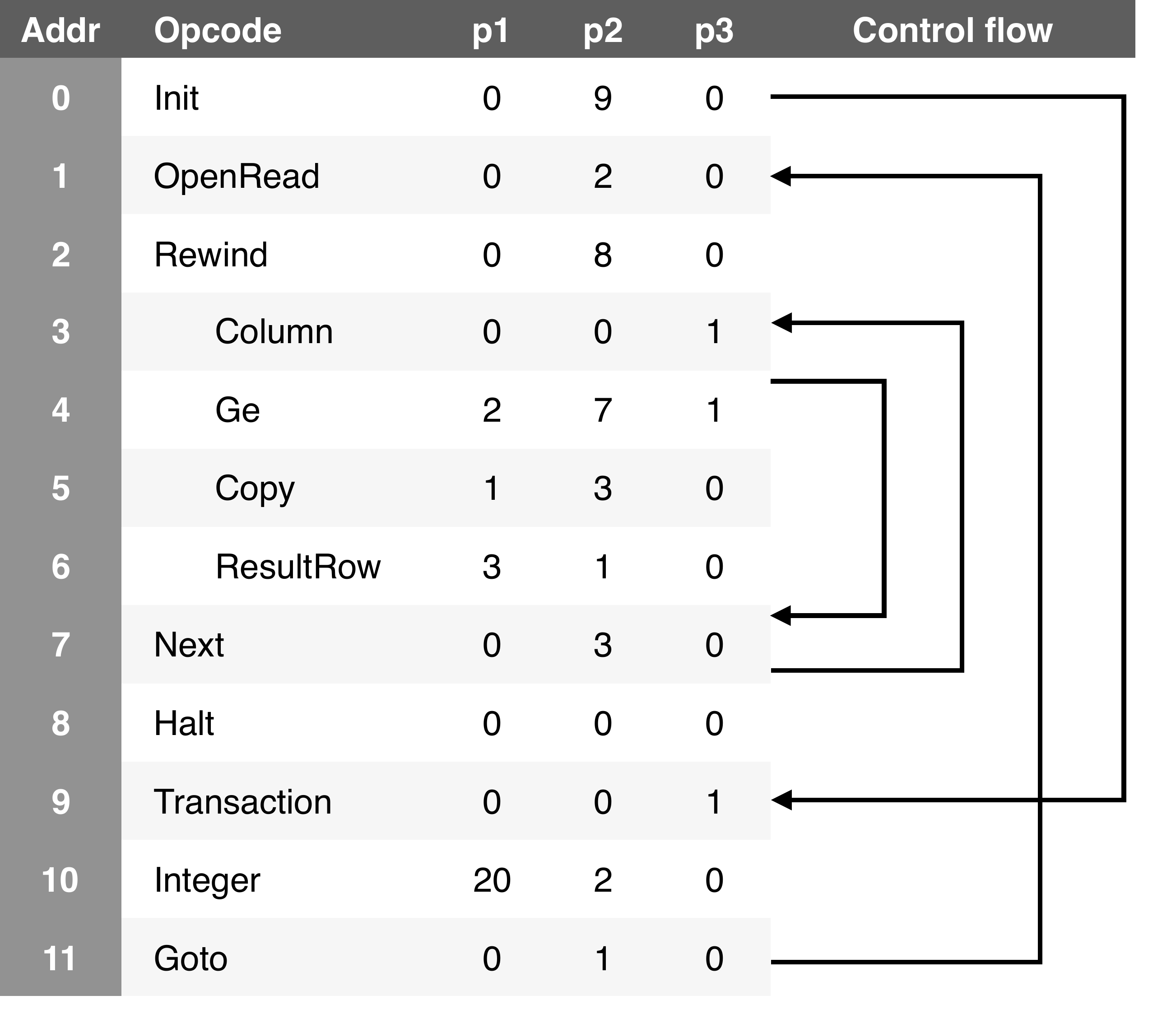}
\caption{A simple query execution plan (abridged)}
\label{execplan}
\end{figure}
To stop the execution, an opcode performs a jump to a label, such as \texttt{vdbe\_return} or \texttt{abort\_due\_to\_error} which handle the release of the resources.

Figure \ref{execplan} illustrates the bytecode for a simple query execution plan. The first instruction sets up some internal values and jumps to the position specified in its \texttt{p2} operand to Line 9. After starting a transaction, an integer with the value from \texttt{p1} (20) is stored in a virtual register at position \texttt{p2} (2). After that, the program counter moves to Line 1, where a new cursor is obtained. The cursor is set to point to the first row at Line 2. Line 3 marks the beginning of the main loop of the query.  \texttt{Column} retrieves the value from \texttt{p2}-th column and stores in \texttt{p3}. \texttt{Ge} jumps to position \texttt{p2} if the value of register \texttt{p1} is greater than or equal to the value of \texttt{p3}. Otherwise, it proceeds to \texttt{Copy}  and to \texttt{ResultRow}, the latter sets the virtual program counter to the correct value to be restored later and returns the return code. The interpretation of the bytecode continues from the last position after the callback function returns. \texttt{Next} moves the table cursor to the next position and moves the execution to the start of the loop on Line 3.

\subsection{Direct threading}
   Many implementations of interpreted programming languages, for instance, CPython, use ``direct threading''. This technique eliminates the \texttt{switch} statement by using the compiler extension first introduced by GCC but now also supported by other compilers known as ``Labels as Values''. This extension allows \texttt{goto} statements to jump to labels that are calculated at runtime. By using this technique, the bounds checking done by the \texttt{switch} statement is removed, which decreases the number of instructions executed per one virtual instruction. Typically, a \texttt{switch} statement is implemented by using a jump table. However, the offset to the jump table is calculated in only one place of the program. It is commonly believed that branch prediction can be improved by calculating the offset to the jump table in each opcode which is how it is done in direct threading \cite{ertl2003structure}. 

\section{JIT ENGINE OVERVIEW}

\begin{figure}
\includegraphics[width=0.45\textwidth]{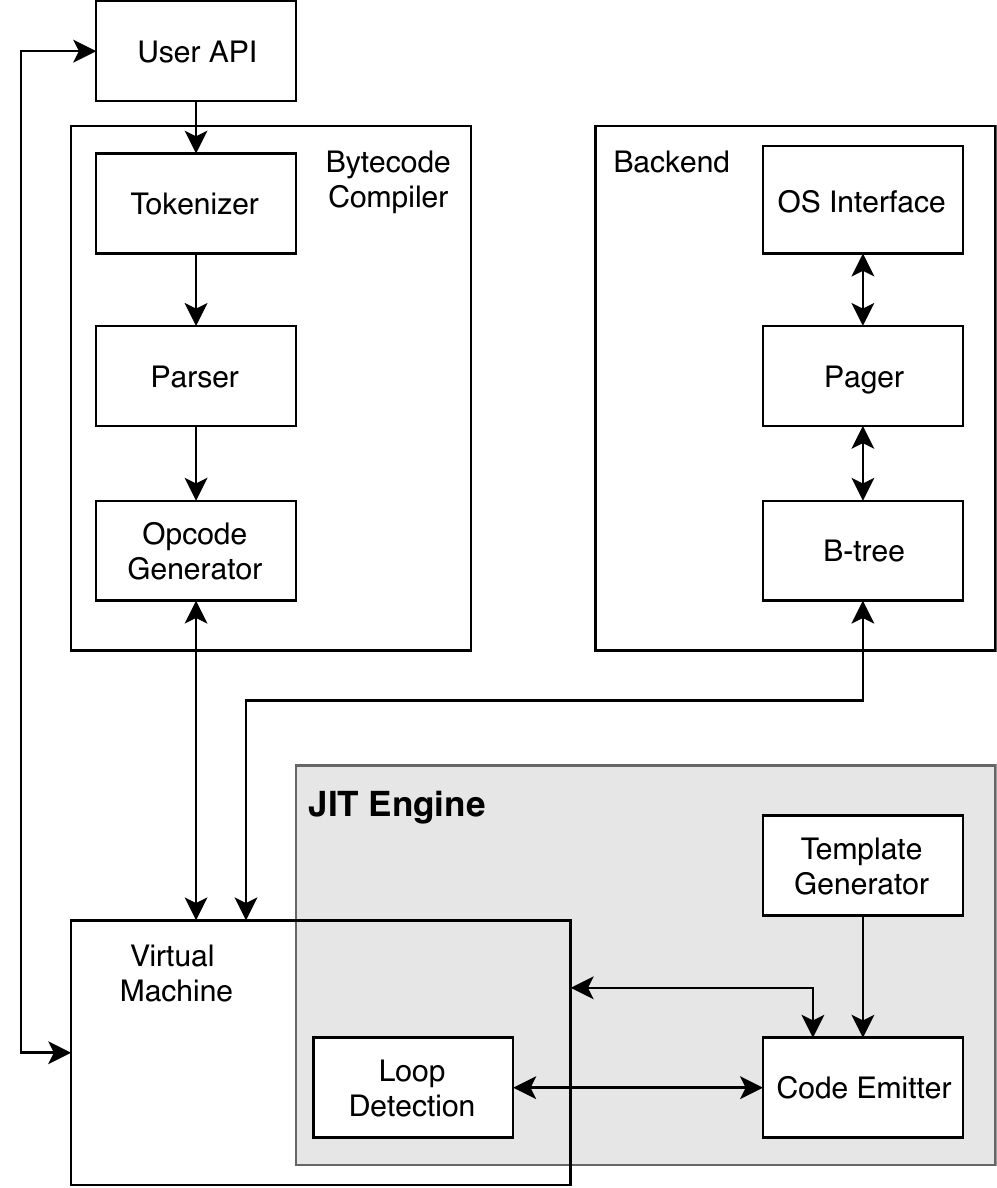}
\caption{System overview}
\label{overview}
\end{figure}

The JIT engine consists of three main components: a loop detector, a template generator, and a code emitter. Figure \ref{overview} provides a top-level overview of the whole system. During build time the template generator extracts C macro templates from the SQLite source code that can be dispatched at runtime by the code emitter. The loop detector keeps track of the instructions executed by the VDBE and invokes the code emitter once a loop has executed enough times. The code emitter composes and compiles a new function implementing the functionality of the loop.

\subsection{Runtime code execution}
During the interpretation of the query execution plan, the JIT engine records the number of times a loop is executed. A loop is detected when the program counter moves to the position in the program which precedes the current one. At that point, a counter contained in the operation is incremented. The position from which the jump happened is also recorded. When a counter becomes greater than the specified threshold, a ``hot loop'' is detected, and the code generation procedure is called. Hot loop detection allows reducing compilation times by only compiling the parts of the query plan that are run repeatedly. The code emitter generates a temporary file containing the definition of the function that will implement the functionality of the loop. The emitter iterates over the instructions in the loop and adds the corresponding macro invocations to the function body. The file is then compiled to a shared object which is linked with the SQLite library. The new object is loaded, and the function pointer is recorded in the operation structure. Now whenever the interpreter reaches the beginning of the hot loop, it calls the compiled function.

The runtime code compilation relies on the existence of macro templates implementing the functionality of the instructions contained in a hot loop. The structure of the interpreter facilitates the creation of such templates. After some modifications, each case block can be converted into an opcode template. However, the number of the required templates make this process tedious which creates a need for automation. Additionally, automating this process allows only to maintain the codebase of the original interpreter. The following section describes how the templates are generated.

\begin{figure}
  \centering
  \begin{subfigure}[b]{0.45\textwidth}
    \begin{lstlisting}[numbers=left,language=C]
case OP_Ge: {
  pIn3 = &aMem[pOp->p3];
  pIn1 = &aMem[pOp->p1]; 
  if(pIn3->u.i >= pIn1->u.i){ 
    goto jump_to_p2;
  }
  break;
}
    \end{lstlisting}
    \caption{A simplified opcode implementation in a case block}

    \label{op_before}
  \end{subfigure}
  \begin{subfigure}[b]{0.45\textwidth}
    \begin{lstlisting}[numbers=left,language=C]
#define GE_TEMPL(pos, next, P1, P3, P2) \
L#pos: { \
  pOp = &aOp[pos]; \
  pIn3 = &aMem[P3]; \
  pIn1 = &aMem[P1]; \
  if(pIn3->u.i >= pIn1->u.i){ \
    goto L#P2; \
  } \
  goto next; \
}
    \end{lstlisting}
    \caption{A generated instruction template}
    \label{op_after}
  \end{subfigure}
  \caption{Transformation of an opcode}
  \label{transform}
\end{figure}

\subsection{Instruction template generation}

To prepare the source code for template generation the file containing the implementation of the VDBE is preprocessed using dummy system library header files, and the \texttt{no inline} compiler directives are removed. The preprocessing reduces the bloat by removing the unnecessary code of system libraries and allows for the file to be efficiently parsed into an abstract syntax tree (AST) representation with pycparser \cite{pycparser}. Next, the case blocks of the opcode implementation being generated are located in the AST by traversing it in the depth-first order. At this point, the first parameter of the macro is created, which is the position of the instruction in the program. The position is used to create a label by which the invocation of the template is addressed.  

SQLite extensively uses \texttt{goto} statements which need to be modified for the future compiled functions to work correctly. The \texttt{goto} statements that are used for error handling are converted into return statements, which moves the burden of resource deallocation to the interpreter. The other crucial type of \texttt{goto} statements in SQLite is \texttt{goto jump\_to\_p2}. The section of the code at that label is responsible for changing the value of the program counter to the appropriate position. The \texttt{jump\_to\_p2} label is therefore replaced by the label of the template invocation corresponding to the instruction at the address contained in the \texttt{p2} register.
The rest of the \texttt{goto} statements are made local to the macro invocation by appending the position of the instruction to the labels. The \texttt{break} statement in the case block is also replaced by a \texttt{goto} statement with the label addressing the template invocation of the next instruction in the sequence. See Fig. \ref{transform} for a simplified example of a conversion. In Fig. \ref{op_before} Line 3 sets the current operation to the correct value, Line 7 contains a \texttt{goto} statement which is completed at code generation time with the value of \texttt{p2}.

Additionally, some of the constants that are known at compile time are replaced and added to the parameters of the macro. The constants are values of the first three virtual registers and the opcode value. The last is of particular interest because it allows for some opcode specialization. In some cases, multiple opcodes are implemented in one case block using the fall through mechanism. The specifics of the implementation are determined based on the value of the opcode. By inserting the value of the opcode at compile time, we can force the compiler to remove the redundant code during compilation.
Finally, a statement which sets the pointer to the current operation structure is added to the beginning of the block.
After the macro has been generated a function which writes the name of the macro to the temporary file with the appropriate parameters is automatically added to the emitter function. 
%See Fig. \ref{emitter} for an example.

% \begin{figure}
% \begin{lstlisting}[numbers=left,language=C]
% static void
% emitGE(Vdbe *p, Op *pOp, TxtBuf **buf, int pos) {
%    writeToBuf(buf, "GE_TEMPL(%d,%d,%d,%d,%d); \n", pos, pos+1, pOp->p1, pOp->p3, pOp->p2);
% }
% \end{lstlisting}
% \caption{Example of a generated emitter function}
% \label{emitter}
% \end{figure}

\section{EXPERIMENTAL STUDY}
The primary goal of the experimental study was to evaluate the potential benefits of the system using JIT compilation and measure the effect of reducing interpreter overhead on query execution time. The number of operations that have to be executed for each tuple heavily influences the interpreter overhead. Therefore, to achieve our goal we designed a microbenchmarking program that can generate a query with an arbitrary number of instructions in a hot loop and measure the relation between CPU execution time and the number of operations in a loop. Experiment A compares the performance of the unmodified version of SQLite, a version that uses JIT compilation and a version that was modified to use a common technique to reduce interpretation overhead that does not involve compilation. Additionally, to assert that JIT compilation is not only beneficial in one particular context, in experiment B we measure the effect of query selectivity on the performance of the systems. For experiment C the opcode implementations were manually specialized for the datatype that is used in the microbenchmark to estimate the potential benefits of automatic opcode specialization.
All tests were conducted on a system with Intel Xeon W-2145 CPU and 128 GB of physical memory running Fedora 26 (kernel 4.14.11). The version of SQLite used is 3.23.0 which was compiled by GCC version 7.3.1 with the \texttt{-O3} flag \footnote{The source code is available on GitHub: \url{https://github.com/AlexKashuba/SQLite_JIT}}. The code generated by JIT was compiled with the \texttt{-O2} flag because it provided the best balance between compilation speed and execution time. A separate process with the SQLite shell was created for every query execution. Each measurement is an average over five runs.

JIT techniques are most useful when the number of operations to interpret grows large. To test how the performance of the systems depends on the number of instructions involved in the execution plan we generate queries with a number of conditions in the \texttt{where} clause. Since SQLite invokes a callback function for each produced result, which may affect performance, we add conditions that cannot change the number of rows produced by the query.  To increase the number of operations in the loop, the test generator adds conjunction of comparisons that will always evaluate to false, such as \texttt{(i < 200 and i > 300)}. The SQLite query optimizer is not able to detect the optimization opportunity and eliminate the comparisons. The pairs of comparisons are connected with disjunction. See an example query in Fig. \ref{query1}. The queries were executed on a dataset consisting of a hundred million random numbers, and the table did not include an index or a primary key to force a full table scan.

\begin{figure}
\includegraphics[width=0.5\textwidth]{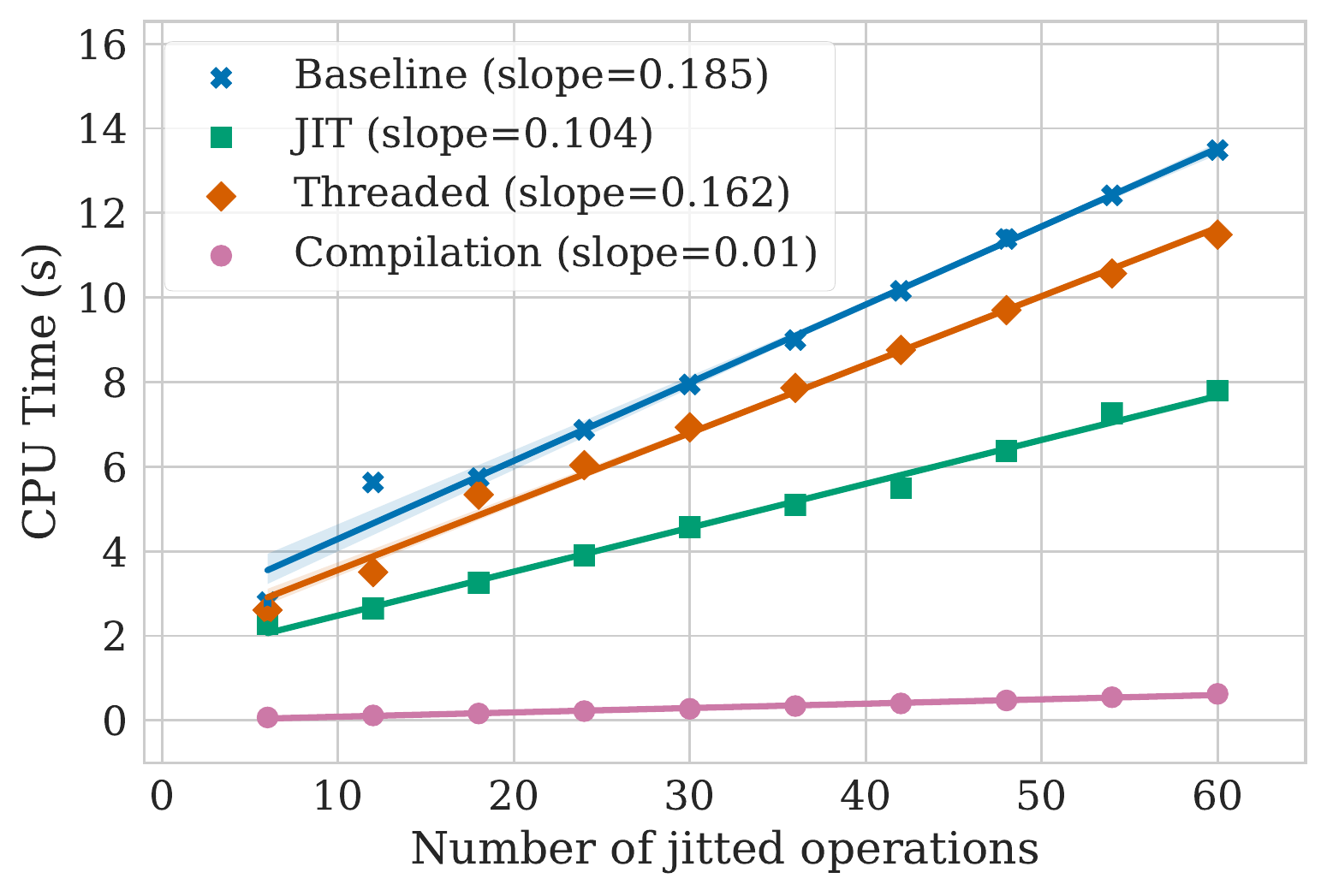}
\caption{Experiment A. Comparison of performance on queries with no results}
\label{test1}
\end{figure}

\begin{figure}
\begin{lstlisting}[language=SQL]
SELECT i FROM test WHERE
(i<1 AND i>6) OR
(i<101 AND i>106) OR
(i<201 AND i>206);
\end{lstlisting}
\caption{Example of a test query}
\label{query1}
\end{figure}

\subsection{Number of operations and system performance}
Experiment A measures the increase in CPU processing time in relation to the number of operations in the main loop of the query plan. The queries included up to sixty operations. For comparison, TPC-H Q5 includes a total of forty-four instructions in its three loops. In order not to measure the overhead associated with materializing the results the queries involved in experiment A produce an empty set. The experiment compares the standard version of SQLite as the primary baseline, the version with the JIT engine, and the second ``threaded'' baseline version.
The third version used direct threading and was created to see how much of the performance improvement could be attributed to eliminating the overhead of the bytecode interpreter compared to other potential effects such as improved cache access patterns.  Direct threading is a technique that eliminates the \texttt{switch} statement by using the compiler extension that allows the use of computed \texttt{goto} statements. It is discussed in further detail in the background section of the paper.  

The results of the evaluation are presented in Figure \ref{test1}.  The JIT version consistently outperforms both baselines with the margin growing with the number of operations reaching the maximum speedup of 1.72x. The time required to compile the function produced by the JIT engine is also displayed. The compilation time does not appear to be a significant slowdown even for complex queries.

Even though it has been argued that modern branch predictors do not require direct threading and can correctly predict the result of the offset calculation used in a \texttt{switch} statement \cite{rohou2015branch}, the new threaded baseline succeeds in reducing the interpreter overhead and demonstrates better performance than the original SQLite.  However, it is unclear whether the improvement should be attributed to the improved branch prediction or other effects.
\begin{figure}
\includegraphics[width=0.5\textwidth]{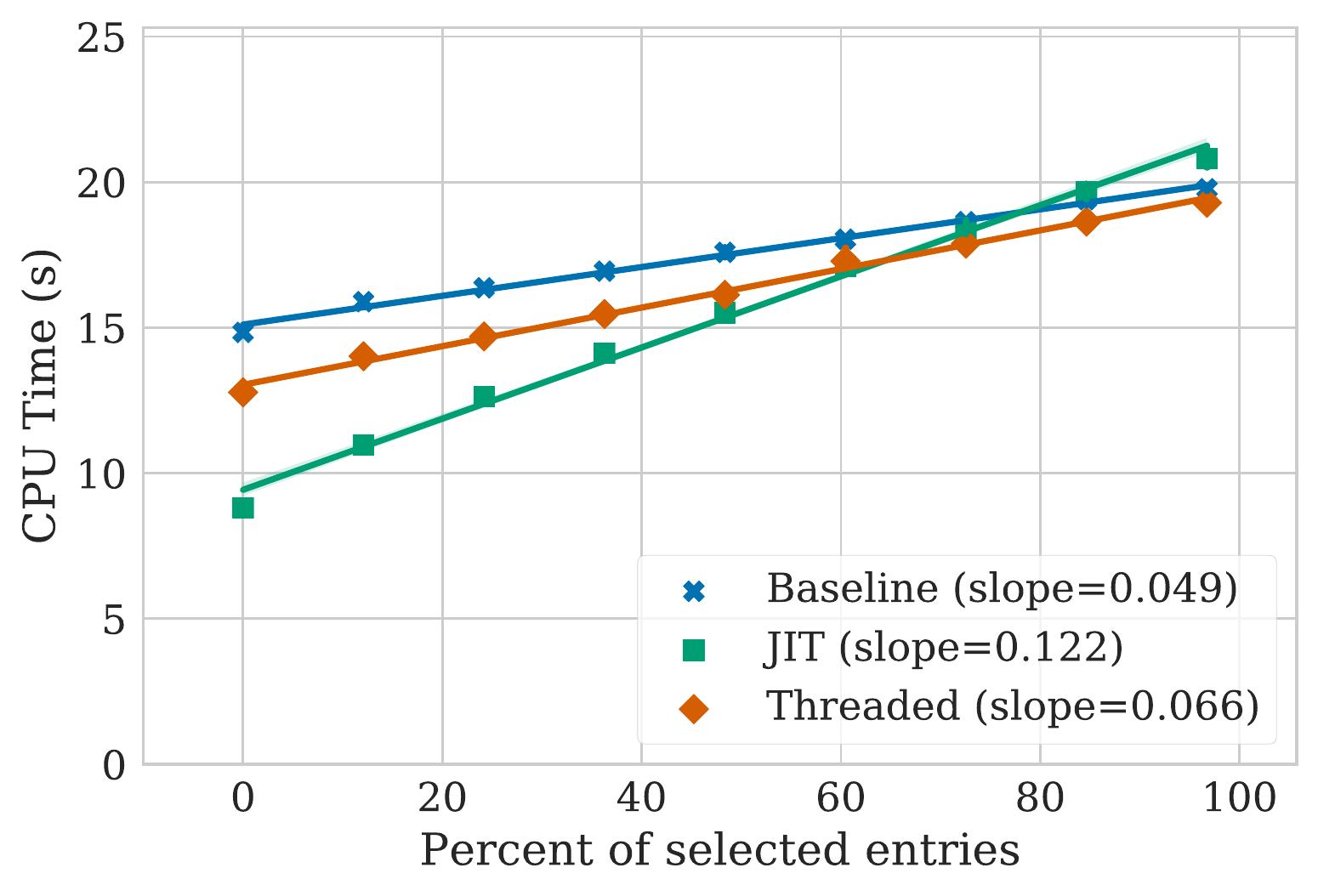}
\caption{Experiment B. Comparison of performance on queries with various degrees of selectivity}
\label{test2}
\end{figure}

\subsection{Effects of query selectivity}
Whenever SQLite finds a tuple satisfying the selection condition, a callback function is executed. Experiment B  evaluates how the number of produced rows affects query execution time to find out whether query selectivity influences the JIT version more than the standard one. The query used in the experiment contains 65 operations in the main loop and is produced by the microbenchmark with an additional condition such as \texttt{i<2000} that evaluates to true in a subset of rows. For each measurement, the newly added condition is tweaked to include more rows and the execution CPU time is recorded. 
The results of the experiment can be seen in Figure \ref{test2}. Query selectivity does affect the JIT version more compared to the baselines. The query processing time in the JIT version grows 2.21 times faster than in the standard one rendering the benefits of JIT negligible when more than sixty percent of the rows are selected. This effect can potentially be explained by the increase in the number of indirect calls to the function implementing the functionality of the hot loop of the query. If that is the case, the overhead of the indirect call could potentially be amortized over multiple rows by buffering the results instead of exiting the JIT function each time a new result is produced.

\begin{figure}
\includegraphics[width=0.5\textwidth]{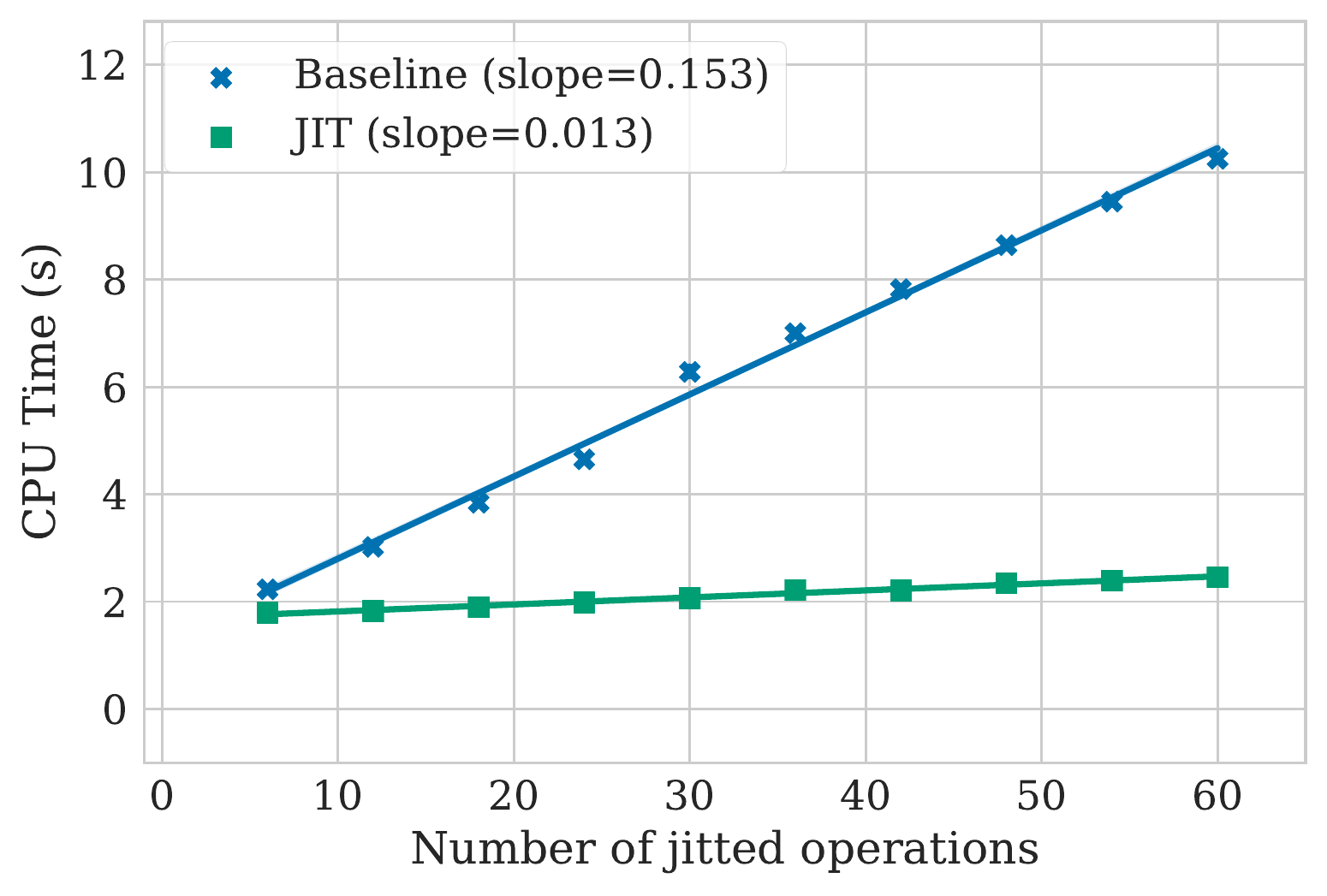}
\caption{Experiment C. Comparison of performance using specialized opcode implementations}
\label{test3}
\end{figure}

\subsection{Manual opcode implementation specialization}
SQLite is a dynamically typed system, and the opcode implementations have to perform many type checks for every tuple to execute their functions, which results in inefficient execution. By using the information available at runtime, such as the table schema, a JIT engine could generate code that is specialized for a particular query.  To investigate the potential benefits of opcode specialization we manually specialized the opcodes used in comparisons to process integer values and added a simple fallback condition that stops the execution in case the type of the attribute is not the same as the expected one. 

Figure \ref{test3} shows the results of the experiment. The JIT version performs remarkably well as the execution time grows extremely slowly with the number of the operations. The new system achieves the maximum speedup of 4.17x over the baseline.

\section{RELATED WORK}
The HIQUE system is one of seminal works in JIT compilation for query execution \cite{krikellas2010generating}. The authors presented a system that translated SQL to C using a collection of templates and exhibited remarkable performance significantly outperforming the competition. However, the query execution engine of HIQUE was created from scratch and did not integrate into an existing system. The templates HIQUE uses were therefore manually written rather than generated from source code.

SQPyte is a partial reimplementation of the SQLite VDBE using the RPython framework, which can derive a JIT compiler from a description of an interpreter \cite{bolz2016making}. This approach to the automatic creation of a JIT engine also allows developing and maintaining only the interpreter. However, on the TPC-H benchmark suite, SQPyte performed only marginally better than stock SQLite providing a 2.2\% improvement.

The authors of \cite{tahboub2018architect} draw on the idea of Futamura projections \cite{futamura1999partial}, i.e., viewing compilation as the specialization of an interpreter, to produce an efficient query compiler. Through the use of  LMS \cite{rompf2012lightweight}, a generative programming framework, a query interpreter is modified to symbolically execute a query plan while also producing C code. However, the authors show that in a single compiler pass this method can only generate efficient code for data-centric style interpreters and involves modifications to the interpreter code for some optimizations.

In \cite{sharygin2017query} the authors also make use of the idea of Futamura projections and propose to separate the development of the algorithmic part of the query execution system from the methods that reduce the interpretation overhead. They develop a query compiler for PostgreSQL, a DBMS which implements Volcano-style query execution. 
The system automatically finds variables and instructions that store or compute  static data in the LLVM IR of the execution engine. The technique allows the system to combine the code of the interpreter with the data available at runtime to automatically produce an interpreter specialized for a particular query.

\section{CONCLUSIONS}
	In this work, we demonstrate an implementation of JIT compilation integrated into SQLite and a method for automatic generation of the templates used for code generation. The experimental study shows the interpreter overhead to be a significant bottleneck in the execution time of complex queries in SQLite. The JIT system substantially improves the performance of query evaluation as compared to the standard version of SQLite as well as a version implementing direct threading. However, due to the design of SQLite where a callback is executed for each produced result, the benefits diminish when the selectivity of a query becomes large. This problem could be remedied in future work by making the JIT system produce results in batches rather than individually. Lastly, the results of manual opcode specialization show a great potential of JIT compilation in the reduction of query execution time. In future work, the template generation could be improved by generating versions of opcodes specialized for a particular type. This would additionally require adding automatic generation and insertion of fallback conditions into the templates. 

\addtolength{\textheight}{-12cm}   % This command serves to balance the column lengths

%%%%%%%%%%%%%%%%%%%%%%%%%%%%%%%%%%%%%%%%%%%%%%%%%%%%%%%%%%%%%%%%%%%%%%%%%%%%%%%%

\input{main.bbl}

\end{document}